\documentclass{llncs}

\usepackage[titletoc,toc,title]{appendix}
\usepackage{float}
\usepackage{placeins}
\usepackage{graphicx}
\usepackage{color}
\usepackage{caption}
\usepackage{pbox}
\usepackage{rotating}
\usepackage{pdflscape}
\usepackage{adjustbox}
\usepackage{amsmath}
\usepackage{breqn}
\usepackage{subfig}
\usepackage{chapterbib}
\usepackage{cite}
\makeatletter
\def\@seccntformat#1{\@ifundefined{#1@cntformat}%
   {\csname the#1\endcsname\quad}  
   {\csname #1@cntformat\endcsname}
}
\let\oldappendix\appendix 
\renewcommand\appendix{%
    \oldappendix
    \newcommand{\section@cntformat}{\appendixname~\thesection\quad}
}
\makeatother

\usepackage{color}
\usepackage[usenames,dvipsnames,svgnames,table]{xcolor}

\usepackage{hyperref}
\hypersetup{
  linkbordercolor=red!80!black,
  citebordercolor=green!80!black,%
  urlcolor={blue!80!black}
}

\usepackage{geometry}
\usepackage{adjustbox}
\usepackage{letltxmacro}
\usepackage{changepage}
\LetLtxMacro{\originalfigure}{\figure}
\LetLtxMacro{\originalendfigure}{\endfigure}
\renewenvironment{figure}[1][htb]%
  {\originalfigure[#1]
   \begin{adjustwidth*}{-.75in}{-.75in}
  }%
  {\end{adjustwidth*}\originalendfigure}

\begin{document}

\title{A nonlinear impact: evidences of causal effects of social media on market prices}
\titlerunning{A nonlinear impact: evidences of causal effects of social media on market prices}  
%
\author{Th\'{a}rsis T. P. Souza\inst{1,}\footnote{Correspondence author: T.Souza@cs.ucl.ac.uk.} \and Tomaso Aste\inst{1,2}}

\authorrunning{Souza et al.} 
%
\tocauthor{Souza et al.}
\institute{Department of Computer Science, UCL, Gower Street, London, WC1E 6BT, UK\and
Systemic Risk Centre, London School of Economics and Political Sciences, London,  WC2A 2AE, UK.}

\maketitle              

\begin{abstract}
Online social networks offer a new way to investigate financial markets' dynamics by enabling the large-scale analysis of investors' collective behavior.
We provide empirical evidence that suggests social media and stock markets have a nonlinear causal relationship.
We take advantage of an extensive data set composed of social media messages related to DJIA index components.
By using information-theoretic measures to cope for possible nonlinear causal coupling between social media and stock markets systems, 
we point out stunning differences in the results with respect to linear coupling. 
Two main conclusions are drawn: First,
social media significant causality on stocks' returns are purely nonlinear in most cases; 
Second, social media dominates the directional coupling with stock market, an effect not observable within linear modeling.
Results also serve as empirical guidance on model adequacy 
in the investigation of sociotechnical and financial systems.
\end{abstract}

\section{Introduction}
\label{Intro}

Investors' decisions are modulated not only by companies' fundamentals but also by personal beliefs, 
peers influence and information generated from news and the Internet. 
Rational and irrational investor's behavior and their relation with the market efficiency hypothesis \cite{JOFI:JOFI518}
have been largely debated in the economics and financial literature \cite{shleifer2000inefficient}. 
However, it was only recently that the availability of vast amounts of data from online systems paved the way 
for the large-scale investigation of investor's collective behavior in financial markets.

Testing for nonlinear dependence is of great importance in financial econometrics due to its implications in model adequacy, market efficiency, and predictability \cite{doi:10.1080/096031096334105}.
Taking social media as a proxy for investor's collective attention over the stock market, 
we provide empirical evidence that characterize social media impact on market prices as nonlinear.

Previous studies have investigated the predictive power of online expressed opinions
and measures of collective attention on market movements .
News are perhaps the most explored source of information, especially after the availability of electronically
transmitted services and machine readable news \cite{tetlock2007giving, tetlock2008more, Tobias:2013, Lillo2012, citeulike:11703961, 
2011arXiv1112.1051M, HestonRanjan:2014, RePEc:2014, mann2011321, Li2014826, Chan2003223, Zhang_tradingstrategies}.
The use of search engines \cite{PreisCurme:2014,Preis5707, citeulike:12299800, mao2014quantifying, 10.1371/journal.pone.0040014,2011arXiv1112.1051M, JOFI:JOFI1679} 
and Wikipedia \cite{wrap54525} are examples of the extension of this investigation to broader types of online systems. 
In addition to that, social media and micro-blogging platforms 
play an increasingly significant role as proxies of collective intelligence  and sentiment of the real world. 
Not only do they mimic real-world peer-to-peer relationships but they also provide a fine-grained real-time information channel 
that include stories, facts and shifts in collective opinion. 
Nonetheless, to what extent this information flood reflects financial dynamics is a relatively novel topic under great debate 
\cite{Bollen20111, mao2014quantifying, Zhang_tradingstrategies, citeulike:13108056, Timm:2014, 
10.1371/journal.pone.0138441, YangTwitter:2014, Sehgal:2007:SSP:1335998.1336036, Antweiler+Frank:04a, 
RePEc:bla:jbfnac:v:41:y:2014:i:7-8:p:791-830, 2011arXiv1112.1051M, DBLP:journals/eswa/NasseriTC15,Dzeroski:2014, Liu:2015:SAP:2774253.2774354, 1507.00784}.

Recent developments have shown the importance of Twitter as an information channel about financial markets.
An example is the U.S. Securities and Exchange Commission allowance of official company's disclosure via Twitter in compliance with Regulation Fair Disclosure \cite{SEC:2013}.
Several research evidences also indicate that Twitter may describe and predict financial dynamics. 
Among the first and most influential works is Bollen et al. (2011) \cite{Bollen20111}, where
the authors used emotion analytics to forecast movements in the DJIA index. 
Later on, in a report for the European Central Bank, the same authors \cite{mao2014quantifying} 
showed that Twitter collective opinion not only has predictive power over stocks' returns but it actually precedes changes in search volume (Google Trends), a known predictor of economic indicators.
Further, Zheludev et al. (2014) \cite{citeulike:13108056} showed that Twitter can contain statistically-significant lead-time information about securities' returns, 
most remarkably over
future prices of the S\&P500 index.
Sprenger et al. (2014) \cite{Timm:2014} proposed a methodology that quantified the impact of Twitter messages 
in the market as well as identified different types of company specific events. 
Subsequent research by Ranco et al. (2015) \cite{10.1371/journal.pone.0138441} 
reinforced these results while analyzing links among Twitter peaks, excess of stocks' returns and the identification of earnings announcements. 

These recent works provide evidence that exogenous information gathered from sociotechnical systems may be useful to describe financial dynamics. 
However, the current body of the literature presents mixed results on the stocks' returns predictability. 
On the one hand, some researches indicate predictability of price movements using News and social media \cite{tetlock2007giving, tetlock2008more, Bollen20111, mao2014quantifying}.
On the other hand, other studies report weak results \cite{10.1371/journal.pone.0138441, citeulike:13108056} suggesting that 
social media analytics have low power when used alone.  
Moreover, the use of ad hoc functional forms and assumptions in different studies makes it difficult to draw 
general conclusions about the nature of the relationship between sociotechnical systems and stock markets.

We take advantage of an information-theoretic framework to study the causality between social media and stock returns in a nonparametric way.
We detect directional and dynamical coupling while not assuming any particular type of interaction between the systems. 
To our knowledge, our results provide the first empirical evidence that suggests social media and stock markets not only have significant lead-time coupling but also are dominated by nonlinear interactions.

\section{Data Analyzed}
Our analysis is conducted on the 30 components of the Dow Jones Industrial Average (DJIA) index, which we monitored during the two-year period from March 31, 2012 to March 31, 2014. 
The choice of these stocks was due to their representativeness of the stock market (see Supporting Information \ref{sec:companies} for the complete list of companies). 
We consider two streams of time series data: (i) market data, which are given at the daily stock price, and (ii) social media data analytics based on 1,767,997 Twitter messages.
Let $P(t)$ be the closing price of an asset at day $t$, as financial variable we consider the stocks' daily log-returns: $R(t) = \log{P(t)} - \log{P(t-1)}$.

We consider Twitter data analytics as a proxy for the collective opinion over a stock. 
As opinion mining \cite{1507.00955} per se is out of scope of this study, we build our analysis on top of Twitter data analytics supplied by PsychSignal.com \cite{PsychSignal}. 
PsychSignal's natural language processing (NLP) 
uses a sophisticated linguistic based approach to sentiment mining that is able to extract and score the nuanced financial language used by traders in online conversations.


%

\begin{figure}[!t]
\centering
\scalebox{0.5}{\includegraphics{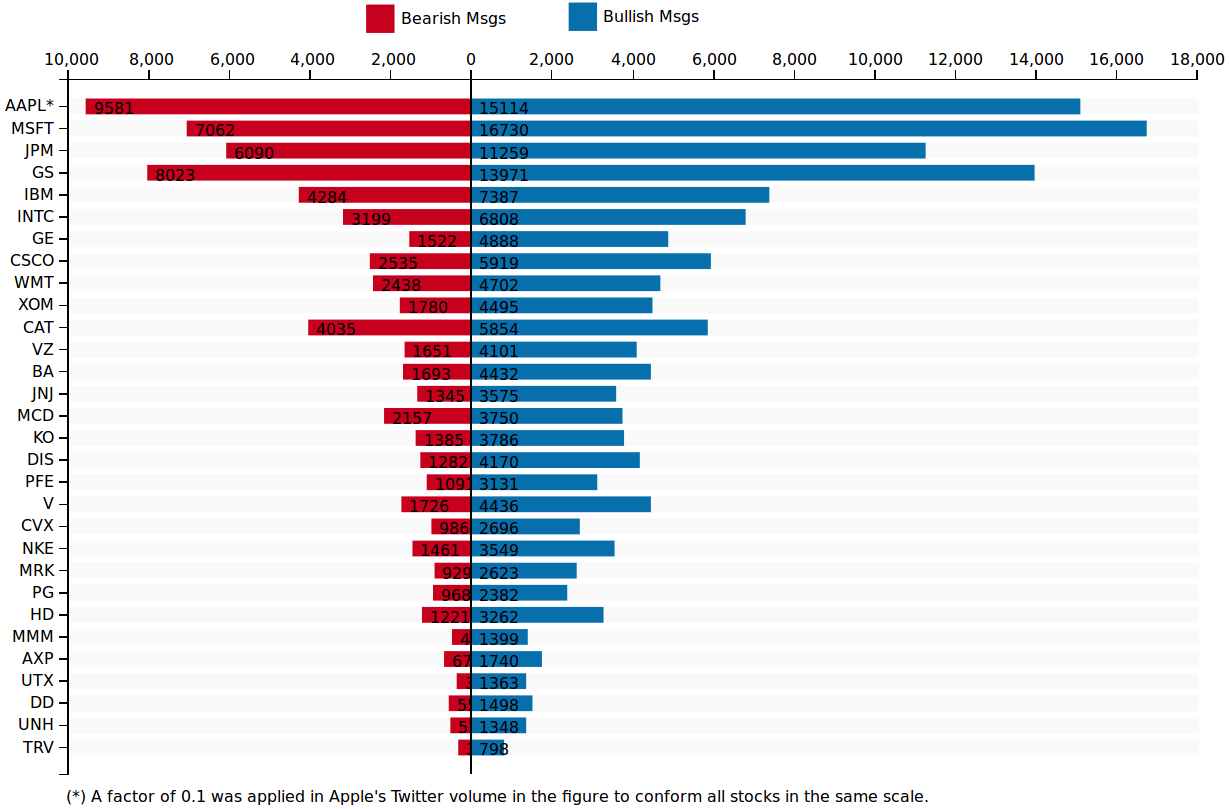}}
\caption{\textbf{Volume of bearish and bullish Twitter messages mentioning a ticker of a stock component of the DJIA index.}
}
\label{fig:companies}
 \end{figure}

We take the daily total number of bullish tweets related to a company as the social media time series $SM(t)$. 
Fig. \ref{fig:companies} shows the volume of bearish and bullish messages for the selected companies.   
A company is defined to be related to a given message if its ticker-id is mentioned as a \textit{cashtag}, i.e., with its name preceded by a dollar symbol, 
e.g., \$CSCO for the company CISCO SYSTEMS INC. In Twitter, a \textit{cashtag} is a standard way to refer to a listed security.
See Supporting Information \ref{sec:Twitter} for further details on the Twitter data analytics.

\section{Social Media and Stocks' Returns: Linear and Nonlinear Causality}
We investigate the characterization of causal inference between social media and stocks' returns under the notion of Granger (G-causality) \cite{Wiener56, granger:econ}.
We test the null hypothesis of social media not causing stocks' returns.
Firstly, we verify this hypothesis with a standard G-causality test under a linear vector-autoregressive framework.
This linear model is tested against misspecification via a BDS test \cite{citeulike:9300127} which is a nonparametric method that is powerful to detect nonlinearity \cite{Barnett97asingle-blind}.
Secondly, we detect significant causalities in possible nonlinear dynamical interactions. 
This is done without assuming any a priori type of interaction.
We consider Transfer Entropy (TE) as the measure for nonparametric causality.
Since its introduction by Schreiber (2000) \cite{PhysRevLett.85.461}, 
TE has been recognized as an important tool in the analysis of causal relationships in nonlinear systems \cite{citeulike:1447442}.
It naturally detects directional and dynamical information \cite{10.1371/journal.pone.0109462}. 
This measure can be interpreted as the information flow between social media and future outcomes of stocks' returns at lag $\Delta t$, controlled by current information on stocks' returns.

Consistent with a nonparametric analysis, we estimate the TE significance via randomized permutation tests. 
If the null hypothesis is rejected, there is evidence of nonlinear causality, otherwise we consider that there is no significant causality.
The hypothesis tests are performed for lags $\Delta t$ ranging from 1 to 10 trading days. 
We apply the Bonferroni correction to reduce the probability of a Type I (false positive) error due to multiple hypotheses testing.

 \begin{figure}[!t]
\centering
\textbf{\Large Social Media $ \rightarrow $ Stocks' Returns}\par\medskip
\scalebox{0.585}{\includegraphics{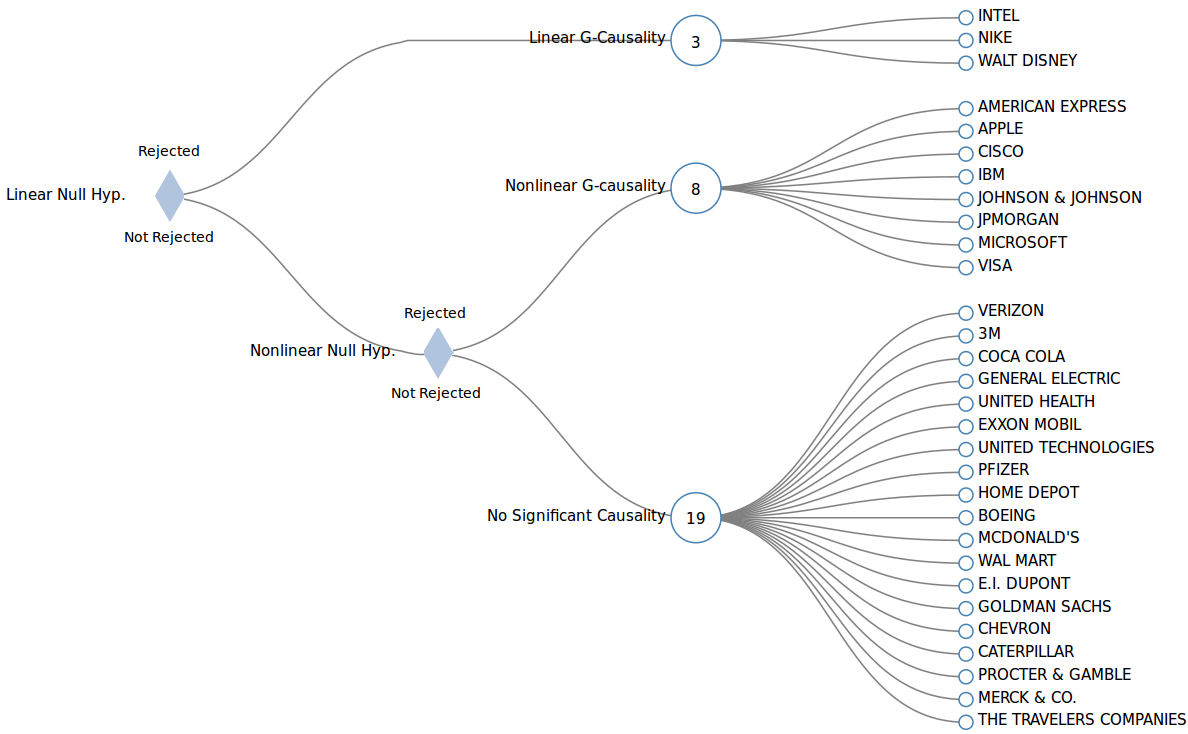}}
\caption{
\textbf{Demonstration that the causality between social media and stocks' returns are mostly nonlinear.}
Linear causality test indicated that social media caused stock's returns only for 3 stocks.
Nonparametric analysis showed that almost 1/3 of the stocks rejected in the linear case have significant nonlinear causality.
In the nonlinear case, Transfer Entropy was used to quantify causal inference between the systems with randomized permutations test for significance estimation. 
In the linear case, a standard linear G-causality test was performed with a F-test under a linear vector-autoregressive framework. 
A significant linear G-causality was accepted if its linear specification was not rejected by the BDS test. 
p-values are adjusted with the Bonferroni correction. Significance is given at p-value $ < 0.05$. 
}
\label{fig:sigpoints-0}
 \end{figure}

 \begin{figure}[!t]
\centering
\scalebox{0.35}{\includegraphics{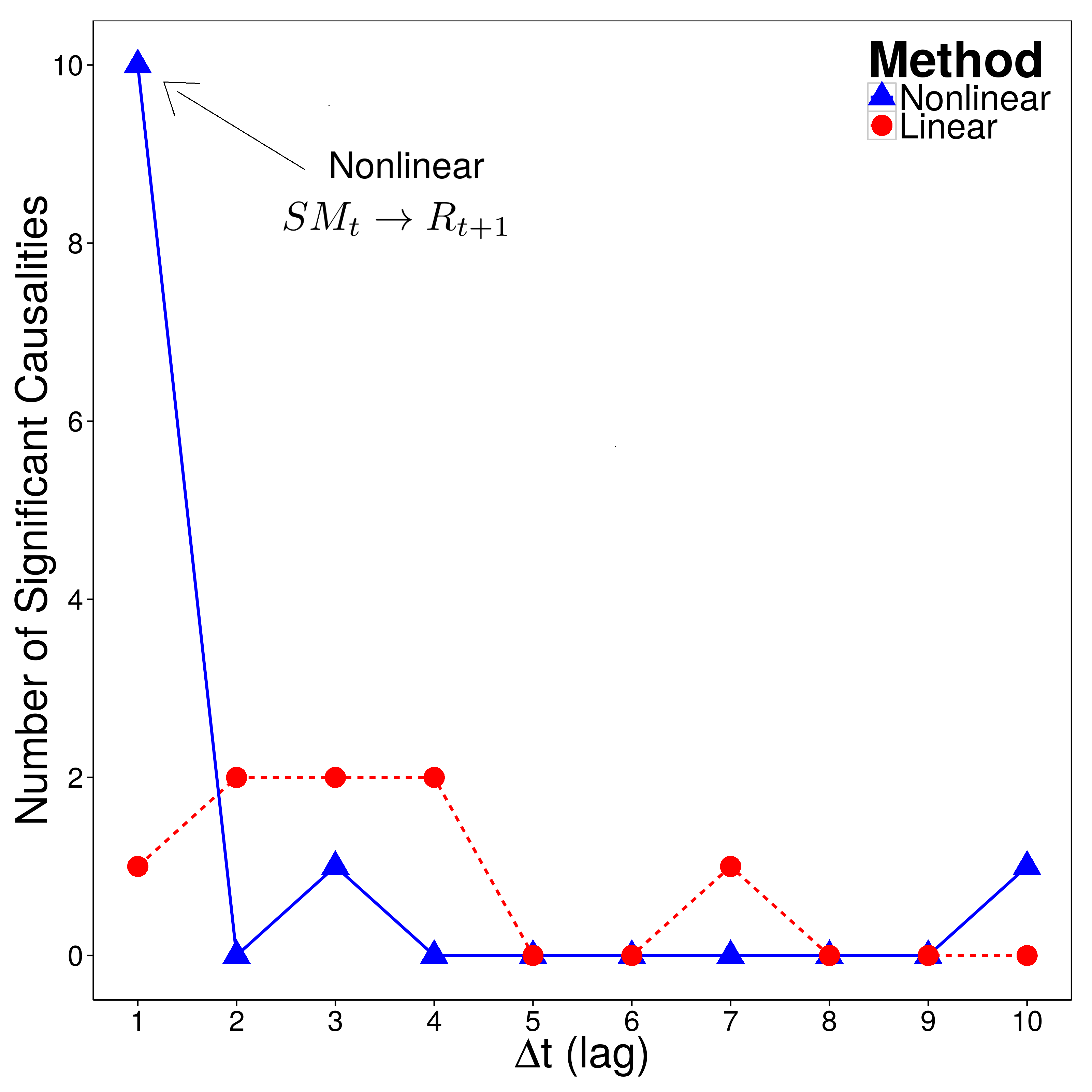}}
\caption{\textbf{Social media causality on stock's return is mostly nonlinear in the next-day period.}
Figure shows the number of companies with significant causality aggregated by lag. 
Causality between social media and next-day stocks' returns presents a stunning difference between linear and nonlinear cases.
Nonlinear analysis identify much higher causality in the first lag. 
Hence, linear-constraints may be neglecting social media causality over stocks' returns, especially in the next-day period.
Further lags present a lower number of significant causalities in both methods.
p-values are adjusted with a Bonferroni correction to reduce the probability of a Type I (false positive) error due to multiple hypotheses testing.
Significance is given at p-value $ < 0.05$.
}
\label{fig:sigpoints2}
 \end{figure}

Fig. \ref{fig:sigpoints-0} shows the significant causality links between social media and stocks' returns considering both cases: nonlinear (TE) and linear G-causality \footnote{See 
Supporting Information \ref{sec:table-causality} for the complete set of p-values obtained.}. 
Linear analysis discovers only three stocks with significant causality: INTEL CORP., NIKE INC. and WALT DISNEY CO. 
Nonlinear analysis discovers that several other
stocks have significant causality. In addition to the 3 stocks identified with significant linear causality, other 8 stocks present purely nonlinear causality.

In Fig. \ref{fig:sigpoints2}, we show the number of stocks with significant causality aggregated by lag of interaction. 
The causality between social media and next-day stocks' returns presents a stunning difference between linear and nonlinear cases. 
From linear G-causality one would say that there is significant causality between social media and
next-day stocks' movements for one stock only. Conversely, nonlinear measures indicated that 10 companies have significant causality in this direction. 
Higher delays show a drop on this number.
These results suggest that linear-constraints are neglecting social media causality over stocks' returns especially in the next-day period or in the short-term.

The low level of causality obtained under linear constraints is inline with results from similar
studies in the literature, where it was found that stocks' returns show weak causality links \cite{Tobias:2013, Antweiler+Frank:04a} and
social media sentiment analytics, at least when taken alone, have very small or no predictive power
\cite{10.1371/journal.pone.0138441} and do not have significant lead-time information about 
stock's movements for the majority of the stocks \cite{citeulike:13108056}.
Contrariwise, results from the nonlinear analyses unveiled a much higher level of causality indicating that linear constraints may be neglecting 
the relationship between social media and stock markets.

 \begin{table}[!t]
\begin{center}
\centering
\caption{\textbf{Nonlinearities found are non-trivial.} Test for linear adequacy for commonly used function forms in the relationship between social media and stocks' returns. Test for misspecification is performed with the BDS test.
$x$ represents the standard linear regression of returns on the social media time series. $\nabla x$ and $\nabla^2 x$ are, respectively, the first and second differencing taken in both time series.
$f(x,vol)$ represents a regression of returns on social media controlled by the stocks' returns daily volatility.
In the log-transformation we apply the function $\log(1+x)$ in both time series. 
The module $|x|$ is applied in the returns time-series which is then regressed over the original social media data.
GARCH(1,1) and ARIMA(1,1,1) transformations were applied on returns, then we regressed the resulting residuals on the original social media time series. See Section \ref{sec:funcs} for the description of the functional forms used.}
\label{tb:BDS}
\begin{tabular}{lcccccccc}
\hline\hline
Ticker&$x$&$\nabla x$ &$\nabla^2 x$ & $f(x,vol)$ & $\log(1+x)$ & $|x|$  & GARCH(1,1) & ARIMA(1,1,1) \tabularnewline
\hline 
CSCO&\scalebox{1.5}{$\circ$}&$ $&\scalebox{1.5}{$\circ$}&\scalebox{1.5}{$\circ$}&\scalebox{1.5}{$\circ$}&\scalebox{1.5}{$\bullet$}&\scalebox{1.5}{$\circ$}&\scalebox{1.5}{$\circ$}\tabularnewline
MSFT&\scalebox{1.5}{$\circ$}&$ $&$ $&\scalebox{1.5}{$\circ$}&\scalebox{1.5}{$\circ$}&\scalebox{1.5}{$\bullet$}&\scalebox{1.5}{$\circ$}&\scalebox{1.5}{$\circ$}\tabularnewline
AXP&\scalebox{1.5}{$\circ$}&$ $&\scalebox{1.5}{$\circ$}&\scalebox{1.5}{$\circ$}&\scalebox{1.5}{$\circ$}&\scalebox{1.5}{$\circ$}&\scalebox{1.5}{$\circ$}&\scalebox{1.5}{$\circ$}\tabularnewline
JPM&\scalebox{1.5}{$\circ$}&$ $&\scalebox{1.5}{$\circ$}&\scalebox{1.5}{$\circ$}&\scalebox{1.5}{$\circ$}&\scalebox{1.5}{$\bullet$}&\scalebox{1.5}{$\circ$}&\scalebox{1.5}{$\circ$}\tabularnewline
IBM&\scalebox{1.5}{$\circ$}&$ $&\scalebox{1.5}{$\circ$}&\scalebox{1.5}{$\circ$}&\scalebox{1.5}{$\circ$}&$ $&\scalebox{1.5}{$\circ$}&$ $\tabularnewline
V&$ $&$ $&\scalebox{1.5}{$\circ$}&$ $&$ $&$ $&$ $&$ $\tabularnewline
JNJ&$ $&$ $&$ $&$ $&$ $&$ $&$ $&$ $\tabularnewline
AAPL&$ $&$ $&$ $&$ $&$ $&$ $&$ $&$ $\tabularnewline
\hline
\multicolumn{9}{c}{\scalebox{1.5}{$\circ$}: Not misspecified; \scalebox{1.5}{$\bullet$}: Not misspecified and with significant G-causality.}
\label{tb:func}
\end{tabular}\end{center}

\end{table}

For the companies identified with nonlinear causality, we tested whether common functional forms and transformations used in the literature can explain the nonlinearities.
We checked model adequacy and causality significance for various functional forms listed in Table \ref{tb:func}, where the results are also reported.
The original linear functional form is adequate for 5 companies but can not explain the nonlinear causality. 
Second-order differencing makes a linear functional adequate for the company VISA, but turns Microsoft as misspecified. 
GARCH and ARIMA filtering were applied in a tentative to separate signal from noise and to linearize the original time series.
Nonetheless, significant causality was not observed.
Other functional forms performed no better than the original linear specification a part from the absolute value transformation.
It is indeed known that social media and news analytics predict absolute changes in market prices \cite{citeulike:13108056, Tobias:2013} better than stock's returns.
This functional form is a proxy for stock returns volatility and therefore it has higher predictability than stock returns. 
Yet, half of the companies still had an unexplained nonlinear causality.

It is clear from the results obtained that the nonlinearities found can not be fully explained by 
returns' volatility neither by naive transformations often employed in related studies. 
This indicates that the nonlinear causality is nontrivial and that there is forecastable structure that can not be explained by commonly-used functional forms.
Therefore, the impact of social media on market prices may be higher than currently reported in related studies, 
because commonly-used functional forms are hiding significant causality, that are here reveled instead with a nonparametric analysis.

\section{Quantifying the Direction of Information Flow}

Transfer-entropy is an asymmetric measure, i.e., $T_{X \rightarrow Y} \neq T_{Y \rightarrow X}$, and thus allows the quantification of directional coupling between systems. 
The Net Information Flow is defined as $\widehat{TE}_{X \rightarrow Y} = TE_{X \rightarrow Y} - TE_{Y \rightarrow X}$. 
One can interpret this quantity as a measure of dominant direction of information flow, i.e., 
a positive result indicates a dominant information flow from $X$ to $Y$ compared to the other direction
or, similarly, it indicates which system provides more predictive information about the other system \cite{Michalowicz:2013:HDE:2601840}.

TE has an intuitive interpretation under the notion of G-causality in the sense that social media may cause (future) stocks' returns only when 
it provides more information to stocks' returns than past stocks' returns themselves. 
In fact, Barnett et al. (2009) \cite{PhysRevLett.103.238701} showed that linear G-causality and Transfer Entropy are equivalent for Gaussian variables.
This result provides a direct mapping between the information-theoretic framework and the linear VAR approach of G-causality.
Hence, it is possible to estimate TE both in its general form and with its equivalent form for linear G-causality. 
The former case is a nonparametric approach that is able to capture possible nonlinear coupling that are likely to be neglected in the latter case.

\begin{figure}[!t]
\centering
\scalebox{0.38}{\includegraphics{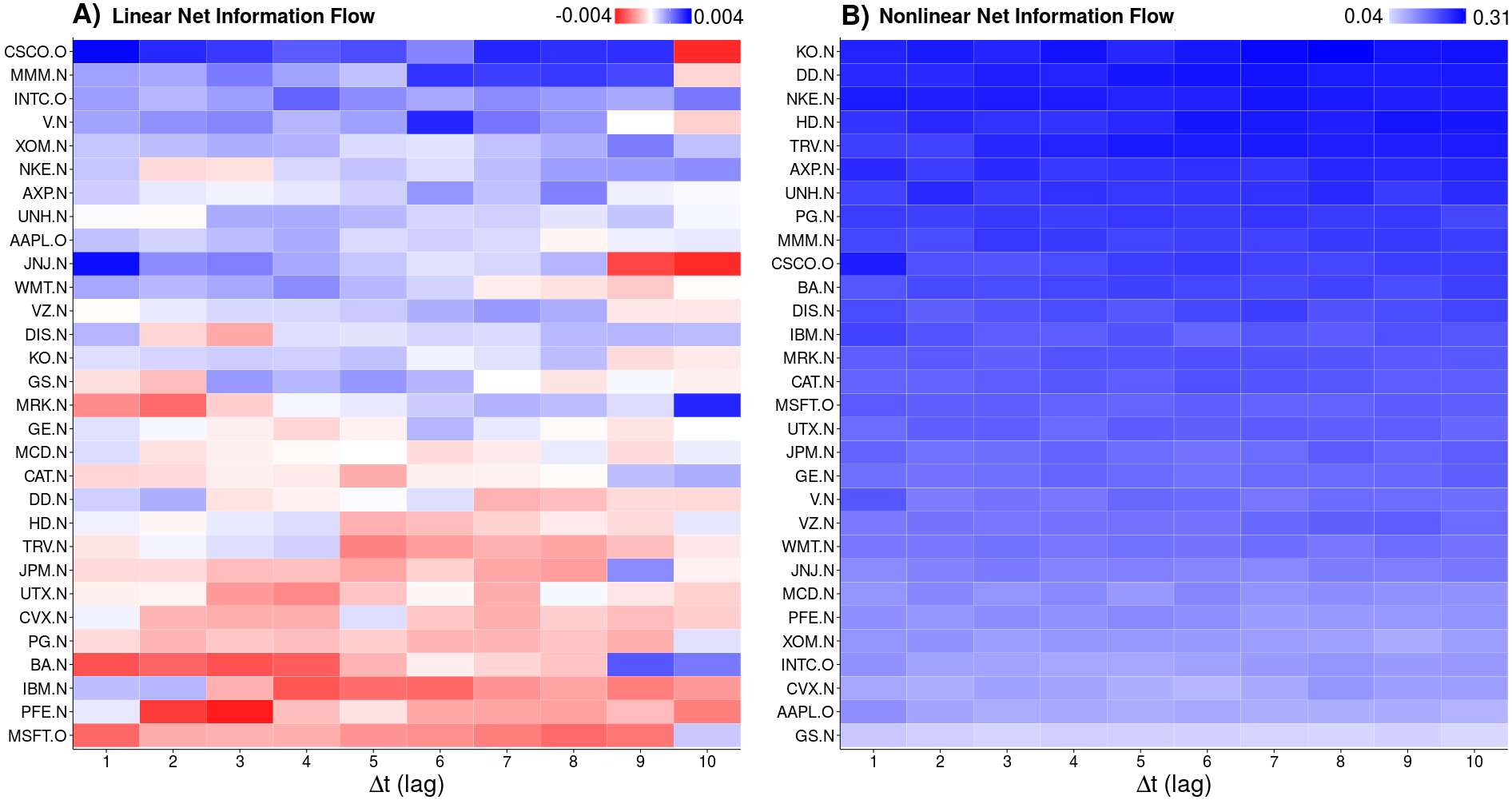}}
\caption{\textbf{Evidence that linear constraints change to a great extent the direction of Information Flow between social media and stock market.}
Figure shows the Net Information Flow from social media to stocks' returns: $\widehat{TE}_{SM \rightarrow R} = TE_{SM \rightarrow R} - TE_{R \rightarrow SM}$.
In A), Net Information Flow is estimated with linear constraints. Positive values indicate that $TE_{SM \rightarrow R} > TE_{R \rightarrow SM}$, 
this is an evidence that information flows from social media to stock returns. 
Contrariwise, negative values indicate that stock market provide more information about social media movements than the opposite.
In B), estimation of Net Information Flow considers nonlinear dynamics. All companies indicate a positive information flow from social media to stocks' returns. 
This indicate that, when nonlinear dynamics are considered, the information flows predominantly from social media to stock market. 
We observe a change of direction of information flow in about half of the companies, compared to the same analysis with linear constraints.
Figure shows the stocks ranked by total Net Information Flow considering all lags, i.e., $\sum_{\Delta t = 1}^{10}{\widehat{TE}_{SM \rightarrow R}}$.
}
\label{fig:TE-NIF}
 \end{figure}

We can therefore quantify the Net Information Flow from social media to stocks' returns using both nonlinear and linear frameworks. 
We investigated which direction of coupling is the strongest and to what extent the consideration of nonlinear dynamics affects the results compared to a linear-constrained analysis.
Fig. \ref{fig:TE-NIF} A) shows the results for the linear case. 
We observe an asymmetry of information, i.e., the systems are not coupled with the same amount of information flow in both directions.
The stocks are clearly divided in two groups of approximately same sizes. One group shows stocks with positive net information flow, 
indicating that social media provides more predictive information about the stock market than the opposite.
A second group of stocks indicates the opposite, i.e., information flows more from stocks' returns to social media than in the other direction. In both cases, the absolute value of net information flow decreases with lag.

Surprisingly, the consideration of nonlinear dynamics unveils a much different scenario. 
Fig \ref{fig:TE-NIF} B) shows the results of the same analysis without linear constraints. 
The net information flow becomes positive for all stocks analyzed. 
This result suggests that social media is the dominant information source indicating
that the information provided by social media contributes more to the description of stock markets dynamics than the opposite.

\section{Summary}
\label{sec:Conclusion}

The present study has revealed that social media has a significant nonlinear impact on stocks' returns.

We analyzed an extensive data set of social media analytics related to stocks components of the DJIA index.
Nonparametric tests for nonlinear specification and causality indicated
three major empirical findings: 
\begin{enumerate}
 \item The consideration of nonlinear dynamics increased the number 
 of stocks with relevant social media signal from 1/10, in the linear case, to more than 1/3 indicating 
 that social media significant causality on stocks' returns are purely nonlinear in most cases;\\
 \item The nonlinearities found were nontrivial and could not be explained by common functional forms used in the literature. 
 This indicates that the impact of social media on stocks' returns may be higher than currently reported in related studies;\\
 \item Nonparametric analysis indicated that social media dominates the directional coupling with stock market; an effect not observable within linear constraints.
\end{enumerate}

We suggest that social media explanatory power on stock markets may be intensified if nonlinear dynamics are considered.
In this respect, we provided strong evidence that supports the use of social media as a valuable source of information about the stock market.

From a methodological point of view, results indicate that a nonparametric approach is highly preferable for the 
investigation of causal relationships between sociotechnical and financial systems.

\section{Methods}
\label{sec:Method}

\subsection{BDS Test for Linear Misspecification}

When applied to the residuals of a linear model, the BDS test \cite{citeulike:9300127} is a powerful test to detect nonlinearity \cite{Barnett97asingle-blind}.
Let $\epsilon_t = (\epsilon_{t=1}, \ldots, \epsilon_{t=n})$ be
the residuals of the linear fitted model and define its $m$-embedding as
$\epsilon_t^m = (\epsilon_{t}, \epsilon_{t-1}, \ldots, \epsilon_{t-m+1})$. The  $m$-embedding correlation integral is given by
\begin{align}
C_{m,n}(\Delta \epsilon) = \frac{2}{k(k-1)}\sum_{s = 1}^{t}{\sum_{t=s}^{n}{ \chi(\| \epsilon_s^m - \epsilon_t^m \|, \Delta \epsilon)    }},
\end{align}
and
\begin{align}
C_{m}(\Delta \epsilon) = \lim_{n\to\infty} C_{m,n}(\Delta \epsilon),
\end{align}
where $\chi$ is an indicator function with $\chi(\| \epsilon_s^m - \epsilon_t^m \|, \Delta \epsilon) = 1$ if 
$\| \epsilon_s^m - \epsilon_t^m \| < \Delta \epsilon$ and zero, otherwise.
The null hypothesis of the BDS test assumes that $\epsilon_t$ is iid. In this case,
\begin{align}
C_{m}(\Delta \epsilon) = C_{1}(\Delta \epsilon)^m.
\end{align}
The BDS statistic is a measure of the extent that this relation holds in the data. It is given by:
\begin{align}
V_{m}(\Delta \epsilon) = \sqrt{n}\frac{C_{m}(\Delta \epsilon) - C_{1}(\Delta \epsilon)^m}{\sigma_m(\Delta \epsilon)},
\end{align}
where $\sigma_m(\Delta \epsilon)$ can be estimated as described in \cite{citeulike:9300127}. 
The null hypothesis of the BDS test indicates that the model tested is not misspecified and it is rejected at 5\% significance level if $\|V_m(\Delta \epsilon)\| > 1.96$.

$\Delta \epsilon$ is commonly set as a factor of the variance ($\sigma_\epsilon$) of $\epsilon$. We report results for $\Delta \epsilon = \sigma_\epsilon/2$ and the embedding dimension $m = 2$. 
We also performed tests for $\Delta \epsilon = \sigma_\epsilon$ and $m = 3$ with no significant differences in the results.

\subsection{Linear G-causality}
Consider the linear vector-autoregressive (VAR) equations:
\begin{align}
R(t) &= {\alpha} + \sum^k_{\Delta t=1}{{\beta}_{\Delta t} R(t-\Delta t)} + \epsilon_t, \label{eq:AR1}\\
R(t) &= \widehat{\alpha} + \sum^k_{\Delta t=1}{{\widehat{\beta}}_{\Delta t} R(t-\Delta t)} +  \sum^k_{\Delta t=1}{{\widehat{\gamma}}_{\Delta t}SM(t-\Delta t)}+ \widehat{\epsilon}_t, \label{eq:AR2}
\end{align}
we test whether $SM$ G-causes $R$ by comparing the errors in the prediction of $R$ in the restricted and unrestricted regression models in Eq. \eqref{eq:AR1} and Eq. \eqref{eq:AR2}, respectively.
Significance estimation is performed via analysis of variance. 
We indicated a significant causality if there is significant causality in at least one of the lags tested.
We adjusted the p-values with a Bonferroni correction to control for multiple hypotheses testing.

\subsection{Functional Forms Tested}
\label{sec:funcs}

Functional forms referenced in Table \ref{tb:func} were used as following.

\subsubsection{Differencing: $\nabla x$.} The first differencing is taken in both social media and returns time series.
\begin{align}
\nabla(R(t), 1) &= \widehat{\alpha} + \sum^k_{\Delta t=1}{{\widehat{\beta}}_{\Delta t} \nabla(R(t-\Delta t),1)} +  \sum^k_{\Delta t=1}{{\widehat{\gamma}}_{\Delta t}\nabla(SM(t-\Delta t),1)}+ \widehat{\epsilon}_t.
\end{align}
The second differencing $\nabla^2 x$ was tested in analogous way.

\subsubsection{$f(x, vol)$.} Represents a regression of returns on social media controlled by the stocks' returns daily volatility.

\begin{align}
R(t) &= \widehat{\alpha} + \sum^k_{\Delta t=1}{{\widehat{\beta}}_{\Delta t} R(t-\Delta t)} +  \sum^k_{\Delta t=1}{{\widehat{\gamma}}_{\Delta t}SM(t-\Delta t)} +  \sum^k_{\Delta t=1}{{\widehat{\theta}}_{\Delta t}vol(t-\Delta t)} + \widehat{\epsilon}_t,
\end{align}
where we consider 
\begin{align}
vol(t) = 2\frac{P_{high}(t) - P_{low}(t)}{P_{high}(t) + P_{low}(t)}
\end{align}
as an approximation of the daily returns volatility. $P_{high}$ and $P_{low}$ are the highest and lowest intraday price value, respectively.

\subsubsection{Log-transformation: $log(x+1)$.}
\begin{align}
\log(R(t)+ 1) &= \widehat{\alpha} + \sum^k_{\Delta t=1}{{\widehat{\beta}}_{\Delta t} \log(R(t-\Delta t)+1)} +  \sum^k_{\Delta t=1}{{\widehat{\gamma}}_{\Delta t}\log(SM(t-\Delta t) + 1)}+ \widehat{\epsilon}_t.
\end{align}

\subsubsection{Absolute value: $|x|$.}
\begin{align}
|R(t)| &= \widehat{\alpha} + \sum^k_{\Delta t=1}{{\widehat{\beta}}_{\Delta t} |R(t-\Delta t)|} +  \sum^k_{\Delta t=1}{{\widehat{\gamma}}_{\Delta t}SM(t-\Delta t)}+ \widehat{\epsilon}_t.
\end{align}

\subsubsection{GARCH(1,1).}

A GARCH filtering was applied in the original returns time series as follows:
\begin{align}
R(t) &= \alpha + \sum^p_{\Delta t=1}{{\beta}_{\Delta t} R(t-\Delta t)} +  \sum^q_{\Delta t=1}{{\gamma}_{\Delta t}\epsilon_{t-\Delta t}}, \label{eq:AR2}
\end{align}
with $p = 1$, $q = 1$ and
\begin{align}
\epsilon_t \sim N(0, R(t)).
\end{align}
The resulting residuals $\epsilon_t$ were then used instead of the original returns time series $R(t)$.

\subsubsection{ARIMA(1,1,1).}
ARIMA filtering was applied in the original returns time series as follows:
\begin{align}
R(t) &= R(t-1) + \alpha(SM(t-1)-SM(t-2))  +  \beta\epsilon_{t-1}.
\end{align}
The resulting residuals $\epsilon_t$ were then used instead of the original returns time series $R(t)$.

\subsection{Nonparametric G-Causality: Transfer Entropy}
\label{Method:TE}
Transfer Entropy (TE) was estimated as a sum of Shannon entropies:
\begin{align}
TE\left(X \rightarrow Y\right) = H\left(Y^P, X^P\right) - H\left(Y^F, Y^P, X^P\right) + H\left(Y^F, Y^P\right) - H\left(Y^P\right),
\end{align}
where $Y^F$ is a forward time-shifted version of $Y$ at lag $\Delta t$ relatively to the contemporaneous time-series $X^P$ and $Y^P$.
We reject the null hypothesis of causality if the Transfer Entropy from social media to stocks' returns is significant.
To remain in a nonparametric framework, the statistical significance of TE was performed using surrogate data.
In that way, 400 replicates of $TE(X_{Shuffled} \rightarrow Y)$ were estimated, where $X_{Shuffled}$ is a random permutation of $X$ relatively to $Y$.
We computed the randomized Transfer Entropy at each permutation for each time-shift ($\Delta t$) from 1 to 10 days. 
We then calculated the frequency at which the observed Transfer Entropy was equal or more extreme 
than the randomized Transfer Entropy of the surrogate data. Statistical significance was given at p-value $< 0.05$.
p-values were also Bonferroni corrected.

The estimation of the empirical probability density distribution, required for the entropy estimation, 
was performed using a Kernel Density Estimation (KDE) method, which has several advantages over the commonly used Histogram based methods (see \ref{sec:KDE}).

\subsection{Net Information Flow}
The Net Information Flow from social media to the stock market is defined as:  $\widehat{TE}_{SM \rightarrow R} = TE_{SM \rightarrow R} - TE_{R \rightarrow SM}$.
For the nonlinear case, transfer entropy was computed as defined in the previous Section \ref{Method:TE}.
Instead, to estimate a linear version of Net Information Flow, we compute Transfer Entropy for the linear case based on the work of \cite{PhysRevLett.103.238701}. 
This work provides a direct mapping between Transfer Entropy and the linear G-causality implemented in the standard VAR framework.
The authors showed that Transfer Entropy and linear G-causality are equivalent for Gaussian variables.

Particularly, assuming the standard measure of linear G-causality for the bivariate case as
\begin{align}
GC_{X \rightarrow Y} = \log\left( \frac{var(\epsilon_t)}{var( \widehat{\epsilon}_t)} \right).
\end{align}
\cite{PhysRevLett.103.238701} shows that:
\begin{align}
GC_{X \rightarrow Y} = 2TE_{X \rightarrow Y},
\end{align}
if all processes ($X$ and $Y$) are jointly Gaussian.

See Supporting Information for the Material (\ref{sec:material}) used and further details in the Transfer Entropy estimation \ref{sec:TE}.

\section*{Acknowledgments}

This work was supported by PsychSignal.com, which provided the social media analytics. T.A. acknowledges support of the UK Economic and Social Research Council (ESRC) in funding the Systemic Risk Centre (ES/K002309/1). T.T.P.S. acknowledges financial support from CNPq - The Brazilian National Council for Scientific and Technological Development.

\bibliographystyle{splncs}
\bibliography{nonlinear}

\newpage

\section*{Supporting Information}
\label{sec:support}
\renewcommand\thesection{SI.\Alph{section}}
\renewcommand\thetable{SI.\Alph{table}}
\setcounter{section}{0}
\setcounter{table}{0}

\section{Material}
\label{sec:material}

\subsection{List of Companies Analyzed}
\label{sec:companies}
The name of the investigated stocks with respective Reuters Instrument Codes (RIC) follow: 
INTEL CORP. (INTC.O), VISA INC. (V.N), NIKE INC. (NKE.N), E.I. DUPONT DE NEMOURS \& CO. (DD.N), JPMORGAN CHASE \& CO. (JPM.N), 
BOEING CO. (BA.N), MERCK \& CO. INC. (MRK.N), PFIZER INC. (PFE.N), MICROSOFT CORP. (MSFT.O), COCA-COLA CO. (KO.N), 
GOLDMAN SACHS GROUP INC. (GS.N), MCDONALD'S CORP. (MCD.N), GENERAL ELECTRIC CO. (GE.N), 3M CO. (MMM.N), UNITED TECHNOLOGIES CORP. (UTX.N), 
VERIZON COMMUNICATIONS INC. (VZ.N), CISCO SYSTEMS INC. (CSCO.O), HOME DEPOT INC. (HD.N), INTERNATIONAL BUSINESS MACHINES CORP. (IBM.N), 
AMERICAN EXPRESS CO. (AXP.N), PROCTER \& GAMBLE CO. (PG.N), APPLE INC. (AAPL.O), UNITEDHEALTH GROUP INC. (UNH.N), CATERPILLAR INC. (CAT.N), 
EXXON MOBIL CORP. (XOM.N), JOHNSON \& JOHNSON (JNJ.N), WAL-MART STORES INC. (WMT.N), WALT DISNEY CO. (DIS.N),  CHEVRON CORP. (CVX.N) and THE TRAVELERS COMPANIES INC. (TRV.N).

\subsection{Twitter Data Analytics}
\label{sec:Twitter}
Twitter data analytics were provided by PsychSignal.com \cite{PsychSignal}. The data are comprised of volume measures and sentiment analytics. 
Twitter messages are classified in two dimensions according to their likelihood of bullishness and bearishness towards a company. 
A company is defined to be related to a given message if its ticker is mentioned. 
The data set are based on English language content and it is agnostic to the country source of the Twitter message. 
The  information is aggregated in a daily fashion and it is composed of the following variables:
\begin{itemize}
  \item symbol: the stock symbol (ticker) for which the sentiment data refer to;\\
  \item timestamp\_utc: date and time of the analyzed data in UTC format;\\
  \item bull\_scored\_messages: this indicator is the total count of bullish sentiment messages;\\
  \item bear\_scored\_messages: this indicator is the total count of bearish sentiment messages.
\end{itemize}
Some messages may be classified as ``neutral'' or at least not having relevant bullish or bearish tones. 
Those types of messages do not affect the bull\_scored\_messages and bear\_scored\_messages scores. 
It is also possible that no messages cite a company in a given day. In that case, the scores are zero.

Table \ref{tb:TA} shows an example of the Twitter sentiment analytics for the company APPLE INC. 
Table \ref{tb:Comp} shows a summary description of the selected companies with the number of bearish/bullish Twitter messages identified in the period.

  \begin{table}[!ht]
\centering
\caption{\textbf{Sample of Twitter Sentiment Analytics for the company APPLE INC}. Twitter messages mentioning a company's ticker (symbol) are classified according to their bullish/bearish tones. 
The table shows a sample of the daily total of bearish and bullish messages classified for the company APPLE INC.}
\label{tb:TA}
\begin{center}
 \begin{tabular}{c c c c} 
\hline
 timestamp\_utc & symbol & bull\_scored\_messages & bear\_scored\_messages \\
 \hline
2015-06-19 & AAPL.O  & 216  & 55 \\
2015-06-20 & AAPL.O  & 66  & 25 \\
2015-06-21  & AAPL.O  & 90  & 24 \\
2015-06-22  & AAPL.O  & 241  & 75 \\ 
2015-06-23  & AAPL.O  & 208  & 75\\ 
2015-06-24  & AAPL.O  & 561  & 211\\ 
2015-06-25  & AAPL.O  & 286  & 107\\ 
2015-06-26  & AAPL.O  & 192  & 82\\ 
2015-06-27  & AAPL.O  & 145 &12\\
2015-06-28  & AAPL.O  & 216  & 69 \\
 \hline
\end{tabular}
\end{center}
 \end{table}  

  \begin{table}[!t]
\centering
\caption{\textbf{Summary table of selected companies.} DJIA stocks along with their total and daily mean number of Bearish and Bullish tweets during the period from March 31, 2012 to March 31, 2014. 
The number of total messages processed include not only messages labeled as bullish and bearish but also neutral messages.}
\label{tb:Comp}
\begin{center}
 \begin{tabular}{l c c c c c} 
\hline
& \multicolumn{2}{c}{Bullish messages} & \multicolumn{2}{c}{Bearish messages} \\
\cline{2-3}  \cline{4-5}
Ticker & Total  & Daily mean & Total & Daily mean & Total Messages\\ \hline
AAPL&151143&279.89&95819&177.443&800638\tabularnewline
MSFT& 16730& 30.98& 7062& 13.078&139343\tabularnewline
JPM& 11259& 20.85& 6090& 11.278& 82265\tabularnewline
GS& 13971& 25.87& 8023& 14.857& 75578\tabularnewline
IBM&  7387& 13.68& 4284&  7.933& 53547\tabularnewline
INTC&  6808& 12.61& 3199&  5.924& 47653\tabularnewline
GE&  4888&  9.05& 1522&  2.819& 41271\tabularnewline
CSCO&  5919& 10.96& 2535&  4.694& 39665\tabularnewline
WMT&  4702&  8.71& 2438&  4.515& 39607\tabularnewline
XOM&  4495&  8.32& 1780&  3.296& 33194\tabularnewline
CAT&  5854& 10.84& 4035&  7.472& 31911\tabularnewline
VZ&  4101&  7.59& 1651&  3.057& 30936\tabularnewline
BA&  4432&  8.21& 1693&  3.135& 30421\tabularnewline
JNJ&  3575&  6.62& 1345&  2.491& 28392\tabularnewline
MCD&  3750&  6.94& 2157&  3.994& 28059\tabularnewline
KO&  3786&  7.01& 1385&  2.565& 26331\tabularnewline
DIS&  4170&  7.72& 1282&  2.374& 25863\tabularnewline
PFE&  3131&  5.80& 1091&  2.020& 24817\tabularnewline
V&  4436&  8.21& 1726&  3.196& 24118\tabularnewline
CVX&  2696&  4.99&  986&  1.826& 21322\tabularnewline
NKE&  3549&  6.57& 1461&  2.706& 20941\tabularnewline
MRK&  2623&  4.86&  929&  1.720& 20708\tabularnewline
PG&  2382&  4.41&  968&  1.793& 20226\tabularnewline
HD&  3262&  6.04& 1221&  2.261& 17550\tabularnewline
MMM&  1399&  2.59&  465&  0.861& 12382\tabularnewline
AXP&  1740&  3.22&  674&  1.248& 12072\tabularnewline
UTX&  1363&  2.55&  369&  0.690& 11255\tabularnewline
DD&  1498&  2.78&  559&  1.037& 10746\tabularnewline
UNH&  1348&  2.50&  532&  0.987&  9196\tabularnewline
TRV&   798&  1.53&  316&  0.604&  7990\tabularnewline
\hline
TOTAL & 287195 & - & 157597 & - & 1767997
\\
\hline
\end{tabular}
\end{center}
 \end{table}

\section{Transfer-Entropy Estimation}
\label{sec:TE}

\subsection{Definitions of information theoretic measures}

Let $X$ be a random variable and $P_X(x)$ its probability density function (pdf). The entropy $H(X)$ is a measure of uncertainty of $X$, and is defined in the discrete case, as:
\begin{equation}
H(X) = -\sum_{x \in X}{P_X(x)\log{P_X(x)}}.
\label{eq:H}
\end{equation}

If the $\log$ is taken to base two, then the unit of $H$ is the \textit{bit} (binary digit). 
Here, we employ the natural logarithm which implies the unit in \textit{nat} (natural unit of information).  

Given a coupled system $(X,Y)$, where $P_Y(y)$ is the pdf of the random variable $Y$ and $P_{X,Y}$ the joint pdf between $X$ and $Y$, the joint entropy is given by:
\begin{equation}
H(X,Y) = -\sum_{x \in X}{\sum_{y \in Y}{P_{X,Y}(x,y)\log{P_{X,Y}(x,y)}}}.
\label{eq:HXY}
\end{equation}
 
The conditional entropy is defined by:
\begin{equation}
H\left(Y\middle\vert X\right) = H(X,Y) - H(X).
\end{equation}
We can interpret $H\left(Y\middle\vert X\right)$ as the uncertainty of $Y$ given a realization of $X$. 

Transfer Entropy can be defined as a difference between conditional entropies: 
\begin{align}
TE\left(X \rightarrow Y\right) = TE\left(X^P,Y^F\middle\vert Y^P\right) =  H\left(Y^F\middle\vert Y^P\right) - H\left(Y^F\middle\vert X^P, Y^P\right),
\end{align}
which can be rewritten as a sum of Shannon entropies:
\begin{align}
TE\left(X \rightarrow Y\right) = H\left(Y^P, X^P\right) - H\left(Y^F, Y^P, X^P\right) + H\left(Y^F, Y^P\right) - H\left(Y^P\right),
\end{align}
where $Y^F$ is a forward time-shifted version of $Y$ at lag $\Delta t$ relatively to the contemporaneous time-series $X^P$ and $Y^P$.

\subsection{Kernel Density Estimation}
\label{sec:KDE}
In the entropy computation, the empirical probability distribution must be estimated.
Histogram based methods and kernel density estimations are  the  two  main methods for that.
Histogram-based is the simplest and most used nonparametric density estimator. 
Nonetheless, it yields density estimates that have discontinuities and vary significantly depending on the bin's size choice.

Also known as Parzen-Rosenblatt window method, the kernel density estimation (KDE) approach approximates the density function at a point $x$ using neighboring observations. 
However, instead of building up the estimate according to bin edges as in histograms, the KDE method uses each point of estimation $x$ as the center of a bin of width $2h$ 
and weight it according to a kernel function. Thereby, the kernel estimate of the probability density function $f(x)$ is defined as
\begin{equation}
\hat{f} = \frac{1}{nh}\sum_{x' \in X}{K\left(\frac{x - x'}{h}\right)}. 
\label{pdf}
\end{equation}

A usual choice for the kernel $K$, which we use here, is the (Gaussian) radial basis function:
\begin{equation}
K(x) = \frac{1}{\sqrt{2\pi}}\exp^{-\frac{1}{2}x^2}. 
\end{equation}

The problem of selecting the bandwidth $h$ in equation \eqref{pdf} is crucial in the density estimation. 
A large $h$ will over-smooth the estimated density and mask the structure of the data. 
On the other hand, a small bandwidth will reduce the bias of the density estimate at the expense of a larger variance in the estimates. 
If we assume that the true distribution is Gaussian and we use a Gaussian kernel, the optimal value of $h$ that minimizes the mean integrated squared error (MISE) is
\begin{equation*}
h^* = 1.06\sigma N^{-1/5},
\end{equation*}
where $N$ is the total number of points and $\sigma$ can be estimated as the sample standard deviation. 
This bandwidth estimation is often called Gaussian approximation or Silverman's rule of thumb for kernel density estimation \cite{silverman}. This is the most common used method and it is here employed. Other common methods are given by Sheather and Jones \cite{sheather1991reliable} and Scott \cite{scott1992multivariate}.

\section{Results of the BDS and causality tests}
\label{sec:table-causality}

 \begin{table}[!t]
\centering
\caption{\textbf{Results of the BDS test. Significance of the null hypothesis of linear specification between Social Media and stocks' returns.} 
Social media bullishness is taken as independent variable with stocks' returns as outcome variable.
p-values higher than 0.05 are evidence of misspecification of the linear functional form indicating a nonlinearly neglected relationship. Lags of up to 10 days were tested.  
p-value $< 0.05$: *; p-value $< 0.01$: **. }
\label{tb:GC-results}
\begin{center}
 \begin{tabular}{|l | c c c c c c c c c c |} 
 \cline{2-11}
 \multicolumn{1}{l|}{} & \multicolumn{10}{c|}{ Lags $\Delta t$ for: $R_t = \sum_{i=1}^{\Delta t}\alpha_i R_{t-\Delta t} + \sum_{i=1}^{\Delta t}\beta_i SM_{t-\Delta t}$} \\
\hline
Ticker & 1 & 2 & 3 & 4 & 5 & 6 & 7 & 8 & 9 & 10\\
\hline
MMM.N&0.928&0.780&0.768&0.999&0.668&0.747&0.739&0.362&0.389&0.676\tabularnewline
AXP.N&0.906&0.987&0.503&0.795&0.734&0.432&0.737&0.336&0.197&0.132\tabularnewline
AAPL.O&0.015*&0.023*&0.028*&0.018*&0.018*&0.008**&0.007**&0.002**&0.003**&0.009**\tabularnewline
BA.N&0.834&0.700&0.591&0.239&0.187&0.363&0.287&0.345&0.187&0.213\tabularnewline
CAT.N&0.001**&0.002**&0.001**&0.000**&0.000**&0.000**&0.000**&0.001**&0.004**&0.002**\tabularnewline
CVX.N&0.113&0.073&0.202&0.330&0.150&0.151&0.389&0.268&0.275&0.439\tabularnewline
CSCO.O&0.404&0.443&0.500&0.463&0.436&0.814&0.883&0.639&0.597&0.590\tabularnewline
KO.N&0.544&0.905&0.712&0.864&0.451&0.734&0.653&0.702&0.471&0.793\tabularnewline
DD.N&0.523&0.373&0.170&0.295&0.759&0.254&0.311&0.199&0.204&0.169\tabularnewline
XOM.N&0.986&0.871&0.504&0.382&0.502&0.704&0.635&0.648&0.178&0.202\tabularnewline
GE.N&0.055&0.016*&0.017*&0.112&0.117&0.129&0.103&0.098&0.101&0.101\tabularnewline
GS.N&0.077&0.010**&0.041*&0.021*&0.028*&0.068&0.102&0.065&0.079&0.014*\tabularnewline
HD.N&0.171&0.092&0.348&0.211&0.079&0.231&0.188&0.034*&0.011*&0.037*\tabularnewline
INTC.O&0.888&0.931&0.808&0.880&0.828&0.858&0.735&0.968&0.756&0.825\tabularnewline
IBM.N&0.184&0.201&0.121&0.126&0.066&0.216&0.393&0.344&0.263&0.209\tabularnewline
JNJ.N&0.000**&0.000**&0.000**&0.000**&0.000**&0.000**&0.000**&0.000**&0.000**&0.000**\tabularnewline
JPM.N&0.171&0.133&0.146&0.305&0.401&0.322&0.270&0.304&0.246&0.798\tabularnewline
MCD.N&0.837&0.970&0.953&0.944&0.497&0.708&0.768&0.775&0.995&0.988\tabularnewline
MRK.N&0.031*&0.026*&0.015*&0.054&0.074&0.134&0.157&0.089&0.126&0.210\tabularnewline
MSFT.O&0.138&0.079&0.129&0.031*&0.009**&0.074&0.066&0.080&0.201&0.136\tabularnewline
NKE.N&0.220&0.126&0.124&0.158&0.055&0.146&0.033*&0.061&0.035*&0.023*\tabularnewline
PFE.N&0.606&0.601&0.599&0.573&0.346&0.405&0.526&0.777&0.651&0.838\tabularnewline
PG.N&0.005**&0.010*&0.012*&0.016*&0.054&0.052&0.015*&0.017*&0.022*&0.017*\tabularnewline
TRV.N&0.000**&0.000**&0.000**&0.000**&0.000**&0.002**&0.003**&0.000**&0.000**&0.000**\tabularnewline
UNH.N&0.007**&0.014*&0.036*&0.016*&0.019*&0.013*&0.006**&0.039*&0.016*&0.011*\tabularnewline
UTX.N&0.176&0.204&0.256&0.577&0.084&0.079&0.185&0.254&0.725&0.585\tabularnewline
VZ.N&0.001**&0.000**&0.000**&0.001**&0.002**&0.000**&0.001**&0.027*&0.000**&0.000**\tabularnewline
V.N&0.001**&0.000**&0.000**&0.000**&0.003**&0.022*&0.049*&0.017*&0.023*&0.023*\tabularnewline
WMT.N&0.200&0.364&0.525&0.937&0.839&0.646&0.719&0.976&0.946&0.753\tabularnewline
DIS.N&0.294&0.310&0.217&0.173&0.421&0.285&0.174&0.179&0.270&0.432\tabularnewline
\hline
\end{tabular}
\end{center}
 \end{table} 

\begin{table}[!t]
 \centering
\rotatebox{90}{
\begin{minipage}{\textheight}
\centering
\caption{\textbf{Significance for nonlinear causality tests}. Lags of up to 10 days were tested 
in both directions of causality: social media causing returns $SM \rightarrow R$ and the opposite, returns causing social media $R \rightarrow SM$.  p-value $< 0.05$: *; p-value $< 0.01$: **. }
\label{tb:TE-results}
\begin{center}
 \begin{tabular}{|l | c c c c c c c c c c | c c c c c c c c c c |} 
 \cline{2-21}
 \multicolumn{1}{l|}{} & \multicolumn{10}{c|}{Lagged $R \rightarrow SM$} & \multicolumn{10}{c|}{Lagged $SM \rightarrow R$} \\
\hline
Ticker & 1 & 2 & 3 & 4 & 5 & 6 & 7 & 8 & 9 & 10 & 1 & 2 & 3 & 4 & 5 & 6 & 7 & 8 & 9 & 10 \\
\hline
MMM.N&0.257&0.248&0.970&0.530&0.198&0.838&0.655&0.767&0.505&0.615&0.410&0.765&0.755&0.570&0.568&0.973&0.980&0.797&0.825&0.945\tabularnewline
AXP.N&0.145&0.275&0.480&0.002**&0.157&0.310&0.380&0.475&0.328&0.142&0.000**&0.297&0.282&0.160&0.407&0.098&0.440&0.152&0.405&0.118\tabularnewline
AAPL.O&0.440&0.995&0.710&0.512&0.912&0.988&0.980&0.897&0.615&0.990&0.000**&0.745&0.590&0.140&0.682&0.585&0.782&0.715&0.095&0.975\tabularnewline
BA.N&0.090&0.475&0.110&0.110&0.718&0.160&0.062&0.345&0.050&0.150&0.287&0.297&0.262&0.070&0.103&0.123&0.095&0.545&0.510&0.130\tabularnewline
CAT.N&0.265&0.562&0.162&0.607&0.170&0.787&0.623&0.235&0.480&0.435&0.055&0.540&0.245&0.228&0.297&0.025*&0.157&0.035*&0.485&0.348\tabularnewline
CVX.N&0.422&0.500&0.710&0.667&0.130&0.040*&0.073&0.955&0.745&0.585&0.270&0.973&0.218&0.382&0.478&0.932&0.528&0.343&0.500&0.330\tabularnewline
CSCO.O&0.083&0.220&0.047*&0.340&0.797&0.605&0.703&0.355&0.325&0.818&0.000**&0.655&0.488&0.528&0.292&0.218&0.500&0.785&0.537&0.645\tabularnewline
KO.N&0.160&0.377&0.095&0.617&0.010*&0.103&0.270&0.405&0.020*&0.073&0.193&0.297&0.365&0.295&0.568&0.162&0.005**&0.015*&0.235&0.575\tabularnewline
DD.N&0.330&0.098&0.055&0.047*&0.758&0.620&0.752&0.722&0.330&0.542&0.535&0.792&0.275&0.295&0.323&0.147&0.478&0.705&0.760&0.377\tabularnewline
XOM.N&0.955&0.677&0.580&0.738&0.838&0.735&0.375&0.593&0.002**&0.375&0.100&0.015*&0.532&0.105&0.662&0.742&0.333&0.722&0.720&0.575\tabularnewline
GE.N&0.698&0.190&0.540&0.262&0.415&0.292&0.660&0.610&0.415&0.377&0.182&0.443&0.633&0.032*&0.260&0.613&0.522&0.277&0.080&0.007**\tabularnewline
GS.N&0.655&0.350&0.083&0.595&0.175&0.252&0.657&0.245&0.580&0.145&0.353&0.735&0.752&0.520&0.200&0.490&0.885&0.722&0.838&0.867\tabularnewline
HD.N&0.345&0.350&0.000**&0.007**&0.080&0.532&0.767&0.490&0.320&0.532&0.495&0.225&0.062&0.200&0.167&0.115&0.392&0.555&0.080&0.177\tabularnewline
INTC.O&0.150&0.902&0.290&0.660&0.088&0.463&0.718&0.910&0.550&0.745&0.000**&0.848&0.248&0.818&0.420&0.570&0.443&0.245&0.330&0.310\tabularnewline
IBM.N&0.037*&0.458&0.248&0.390&0.672&0.205&0.318&0.505&0.463&0.800&0.000**&0.017*&0.333&0.722&0.198&0.637&0.088&0.607&0.052&0.498\tabularnewline
JNJ.N&0.490&0.292&0.275&0.677&0.703&0.282&0.402&0.930&0.508&0.287&0.108&0.015*&0.000**&0.380&0.020*&0.125&0.562&0.290&0.137&0.015*\tabularnewline
JPM.N&0.040*&0.057&0.060&0.353&0.035*&0.017*&0.057&0.262&0.330&0.125&0.000**&0.027*&0.030*&0.010*&0.010*&0.145&0.050&0.012*&0.123&0.000**\tabularnewline
MCD.N&0.088&0.943&0.287&0.885&0.037*&0.680&0.287&0.597&0.848&0.780&0.180&0.502&0.740&0.552&0.565&0.090&0.657&0.392&0.840&0.838\tabularnewline
MRK.N&0.220&0.675&0.345&0.475&0.595&0.767&0.445&0.132&0.213&0.277&0.307&0.593&0.835&0.518&0.568&0.502&0.532&0.190&0.542&0.653\tabularnewline
MSFT.O&0.040*&0.732&0.690&0.200&0.093&0.118&0.162&0.270&0.573&0.390&0.000**&0.468&0.568&0.580&0.302&0.115&0.498&0.407&0.323&0.815\tabularnewline
NKE.N&0.032*&0.287&0.435&0.145&0.152&0.020*&0.243&0.483&0.235&0.147&0.000**&0.167&0.390&0.167&0.627&0.110&0.225&0.440&0.593&0.358\tabularnewline
PFE.N&0.113&0.057&0.762&0.438&0.758&0.167&0.020*&0.762&0.667&0.680&0.017*&0.267&0.290&0.307&0.198&0.110&0.570&0.993&0.873&0.643\tabularnewline
PG.N&0.797&0.387&0.838&0.175&0.645&0.267&0.340&0.118&0.453&0.103&0.215&0.277&0.167&0.297&0.550&0.427&0.200&0.375&0.830&0.988\tabularnewline
TRV.N&0.415&0.040*&0.695&0.330&0.177&0.045*&0.427&0.083&0.395&0.395&0.380&0.238&0.270&0.377&0.115&0.115&0.085&0.132&0.255&0.545\tabularnewline
UNH.N&0.022*&0.035*&0.002**&0.093&0.037*&0.090&0.015*&0.170&0.123&0.245&0.022*&0.030*&0.050&0.050&0.243&0.213&0.100&0.165&0.402&0.152\tabularnewline
UTX.N&0.432&0.528&0.560&0.218&0.280&0.867&0.855&0.228&0.480&0.703&0.713&0.302&0.275&0.848&0.037*&0.682&0.407&0.050&0.325&0.805\tabularnewline
VZ.N&0.098&0.650&0.052&0.090&0.108&0.135&0.463&0.748&0.127&0.600&0.090&0.758&0.535&0.782&0.300&0.630&0.198&0.027*&0.007**&0.373\tabularnewline
V.N&0.400&0.645&0.782&0.353&0.742&0.603&0.218&0.613&0.818&0.915&0.000**&0.970&0.780&0.400&0.130&0.368&0.802&0.483&0.715&0.905\tabularnewline
WMT.N&0.458&0.762&0.787&0.493&0.693&0.262&0.848&0.365&0.365&0.272&0.115&0.382&0.150&0.302&0.267&0.195&0.203&0.445&0.147&0.438\tabularnewline
DIS.N&0.510&0.083&0.448&0.630&0.090&0.640&0.985&0.147&0.262&0.795&0.002**&0.742&0.902&0.550&0.762&0.532&0.130&0.935&0.627&0.665\tabularnewline
\hline
\end{tabular}
\end{center}
\end{minipage}
}
 \end{table}

\begin{table}[!t]
 \centering
\rotatebox{90}{
\begin{minipage}{\textheight}
\centering
\caption{\textbf{Significance for linear causality tests.} Lags of up to 10 days were tested 
in both directions of causality: social media causing returns $SM \rightarrow R$ and the opposite, returns causing social media $R \rightarrow SM$.  p-value $< 0.05$: *; p-value $< 0.01$: **. }
\label{tb:GC-results}
\begin{center}
 \begin{tabular}{|l | c c c c c c c c c c | c c c c c c c c c c |} 
 \cline{2-21}
 \multicolumn{1}{l|}{} & \multicolumn{10}{c|}{Lagged $R \rightarrow SM$} & \multicolumn{10}{c|}{Lagged $SM \rightarrow R$} \\
\hline
Ticker & 1 & 2 & 3 & 4 & 5 & 6 & 7 & 8 & 9 & 10 & 1 & 2 & 3 & 4 & 5 & 6 & 7 & 8 & 9 & 10 \\
\hline
MMM.N&0.024*&0.080&0.131&0.213&0.260&0.292&0.308&0.180&0.232&0.240&0.063&0.122&0.129&0.175&0.029*&0.058&0.077&0.131&0.165&0.238\tabularnewline
AXP.N&0.584&0.687&0.785&0.812&0.905&0.942&0.971&0.987&0.972&0.937&0.671&0.878&0.241&0.341&0.441&0.398&0.425&0.260&0.386&0.158\tabularnewline
AAPL.O&0.068&0.029*&0.066&0.084&0.189&0.073&0.173&0.214&0.284&0.355&0.242&0.434&0.607&0.451&0.527&0.028*&0.039*&0.049*&0.035*&0.026*\tabularnewline
BA.N&0.554&0.813&0.943&0.982&0.852&0.878&0.809&0.645&0.667&0.613&0.370&0.574&0.615&0.382&0.469&0.518&0.505&0.518&0.646&0.430\tabularnewline
CAT.N&0.141&0.240&0.160&0.264&0.269&0.201&0.302&0.231&0.177&0.232&0.967&0.989&0.991&0.995&0.993&0.994&0.967&0.542&0.609&0.687\tabularnewline
CVX.N&0.517&0.658&0.778&0.888&0.294&0.364&0.258&0.253&0.314&0.336&0.678&0.919&0.906&0.965&0.941&0.973&0.986&0.991&0.995&0.997\tabularnewline
CSCO.O&0.870&0.954&0.985&0.984&0.959&0.942&0.968&0.453&0.548&0.606&0.834&0.886&0.945&0.980&0.982&0.988&0.979&0.991&0.996&0.995\tabularnewline
KO.N&0.753&0.562&0.694&0.776&0.477&0.565&0.687&0.658&0.119&0.211&0.149&0.264&0.024*&0.046*&0.084&0.125&0.198&0.009**&0.012*&0.025*\tabularnewline
DD.N&0.283&0.390&0.533&0.088&0.165&0.232&0.332&0.323&0.358&0.419&0.457&0.575&0.759&0.660&0.673&0.557&0.647&0.720&0.809&0.831\tabularnewline
XOM.N&0.640&0.282&0.393&0.549&0.650&0.702&0.866&0.927&0.153&0.217&0.192&0.266&0.375&0.112&0.089&0.084&0.111&0.069&0.062&0.096\tabularnewline
GE.N&0.465&0.327&0.615&0.528&0.574&0.468&0.602&0.685&0.425&0.531&0.719&0.883&0.954&0.745&0.806&0.887&0.735&0.775&0.755&0.625\tabularnewline
GS.N&0.516&0.566&0.822&0.829&0.529&0.517&0.651&0.602&0.628&0.517&0.245&0.287&0.493&0.632&0.708&0.725&0.823&0.831&0.836&0.815\tabularnewline
HD.N&0.293&0.461&0.012*&0.031*&0.047*&0.087&0.134&0.153&0.204&0.226&0.773&0.971&0.473&0.269&0.315&0.431&0.641&0.414&0.427&0.463\tabularnewline
INTC.O&0.040*&0.089&0.130&0.196&0.280&0.352&0.479&0.507&0.329&0.354&0.000**&0.000**&0.002**&0.004**&0.009**&0.019*&0.012*&0.023*&0.037*&0.079\tabularnewline
IBM.N&0.612&0.338&0.258&0.364&0.529&0.345&0.285&0.314&0.300&0.208&0.008**&0.032*&0.073&0.163&0.217&0.128&0.164&0.230&0.264&0.323\tabularnewline
JNJ.N&0.278&0.306&0.538&0.484&0.388&0.501&0.532&0.657&0.739&0.754&0.085&0.224&0.244&0.425&0.563&0.189&0.203&0.126&0.097&0.147\tabularnewline
JPM.N&0.134&0.010**&0.020*&0.042*&0.078&0.092&0.135&0.176&0.245&0.317&0.010*&0.038*&0.053&0.013*&0.041*&0.077&0.070&0.106&0.151&0.104\tabularnewline
MCD.N&0.691&0.542&0.806&0.912&0.936&0.727&0.627&0.694&0.683&0.714&0.176&0.350&0.320&0.529&0.304&0.318&0.250&0.443&0.392&0.482\tabularnewline
MRK.N&0.893&0.232&0.314&0.417&0.144&0.167&0.198&0.178&0.103&0.062&0.864&0.965&0.990&0.995&0.990&0.772&0.679&0.863&0.918&0.932\tabularnewline
MSFT.O&0.245&0.466&0.666&0.616&0.771&0.211&0.225&0.313&0.319&0.411&0.328&0.347&0.451&0.311&0.406&0.124&0.218&0.112&0.173&0.206\tabularnewline
NKE.N&0.054&0.043*&0.028*&0.063&0.104&0.147&0.174&0.180&0.232&0.308&0.008**&0.001**&0.002**&0.003**&0.006**&0.013*&0.017*&0.034*&0.038*&0.050\tabularnewline
PFE.N&0.827&0.974&0.968&0.974&0.965&0.957&0.968&0.949&0.695&0.700&0.821&0.972&0.789&0.880&0.887&0.896&0.904&0.927&0.932&0.963\tabularnewline
PG.N&0.115&0.236&0.235&0.343&0.409&0.475&0.382&0.463&0.599&0.616&0.668&0.476&0.615&0.552&0.527&0.652&0.749&0.766&0.745&0.825\tabularnewline
TRV.N&0.097&0.217&0.262&0.192&0.060&0.036*&0.067&0.014*&0.018*&0.029*&0.532&0.787&0.155&0.229&0.341&0.203&0.284&0.213&0.058&0.081\tabularnewline
UNH.N&0.038*&0.093&0.139&0.226&0.102&0.151&0.173&0.250&0.277&0.322&0.116&0.279&0.423&0.583&0.668&0.731&0.779&0.790&0.846&0.912\tabularnewline
UTX.N&0.700&0.780&0.973&0.988&0.525&0.486&0.432&0.581&0.635&0.566&0.618&0.761&0.249&0.343&0.402&0.507&0.505&0.706&0.347&0.441\tabularnewline
VZ.N&0.348&0.454&0.635&0.808&0.740&0.494&0.456&0.564&0.664&0.748&0.231&0.450&0.586&0.702&0.553&0.434&0.424&0.474&0.026*&0.040*\tabularnewline
V.N&0.033*&0.059&0.097&0.209&0.245&0.269&0.222&0.062&0.095&0.130&0.242&0.369&0.204&0.174&0.069&0.113&0.178&0.158&0.141&0.160\tabularnewline
WMT.N&0.845&0.993&0.957&0.792&0.921&0.894&0.932&0.856&0.899&0.946&0.147&0.126&0.199&0.178&0.508&0.378&0.534&0.489&0.576&0.674\tabularnewline
DIS.N&0.838&0.314&0.513&0.364&0.397&0.519&0.512&0.598&0.656&0.609&0.018*&0.063&0.116&0.212&0.183&0.221&0.005**&0.007**&0.017*&0.024*\tabularnewline
\hline
\end{tabular}
\end{center}
\end{minipage}
}
 \end{table}

\end{document}